# High-efficiency edge couplers enabled by vertically tapering on lithium-niobate photonic chips

Di Jia,[1, †] Qiang Luo,[1, †] Chen Yang,[2] Rui Ma[1], Xuanyi Yu,[1] Feng Gao,[1] Qifan Yang,[2,3] Fang Bo,[1, *] Guoquan Zhang,[1, 4] and Jingjun Xu,[1, 5]

[1]*MOE Key Laboratory of Weak-Light Nonlinear Photonics, TEDA Institute of Applied Physics and School of Physics, Nankai University, Tianjin 300457, China*
[2]*State Key Laboratory for Mesoscopic Physics and Frontiers, School of Physics, Science Center for Nano-optoelectronics, Peking University, Beijing 100871, China*
[3]*e-mail: leonardoyoung@pku.edu.cn*
[4]*e-mail: zhanggq@nankai.edu.cn*
[5]*e-mail: jjxu@nankai.edu.cn*
[†]*These authors contribute equally to this work*
[*]*Corresponding author: bofang@nankai.edu.cn;*

In the past decade, photonic integrated circuits (PICs) based on thin-film lithium niobate (TFLN) have advanced in various fields, including optical communication, nonlinear photonics, and quantum optics. A critical component is an efficient edge coupler connecting PICs to light sources or detectors. Here, we propose an innovative edge coupler design with a wedge-shaped TFLN waveguide and a silicon oxynitride (SiON) cladding. Experimental results show that the coupling loss between the TFLN PIC and a 3-μm mode field diameter (MFD) lensed fiber is low at 1.52 dB/facet, with the potential for improvement to 0.43 dB/facet theoretically. The coupling loss between the edge coupler and a UHNA7 fiber with an MFD of 3.2 μm is reduced to 0.92 dB/facet. This design maintains robust fabrication and alignment tolerance. Importantly, the minimum linewidth of the TFLN waveguide of the coupler (600 nm) can be easily achieved using foundry-available i-line stepper lithography. This work benefits the development of TFLN integrated platforms, such as on-chip electro-optic modulators, frequency comb generation, and quantum sensors.

https://doi.org/ XXXXX.

## 1. INTRODUCTION

In recent years, the field of PICs has witnessed remarkable progress, offering substantial contributions to various applications, including modern telecommunications, sensing and computation. A pivotal factor in these advancements has been the utilization of TFLN [1-4], made possible by its outstanding properties [5, 6] and developed fabrication technologies [7-10]. Leveraging TFLN as a foundation, a series of high-performance photonic devices has been developed, such as the high-efficiency frequency converter [11-17], broadband microcavity frequency comb [18-20], compact amplifiers and lasers [21-23], and high-speed electro-optic modulators [24, 25]. These achievements underscore the pivotal role of TFLN in shaping the landscape of modern photonic technology.

Efficient energy exchange between fiber components and TFLN waveguide is essential for applications of TFLN PICs [26]. Grating couplers and edge couplers are two common devices used for coupling light into and out of on-chip devices. Generally, grating couplers feature large alignment tolerances and are compact and suitable for densely packed photonic circuits [27, 28]. However, the limited coupling efficiency and the polarization sensitivity limits its application scenarios [29]. By contrast, edge couplers have negligible polarization dependence and work in a wide bandwidth [30].

Recently, several edge couplers have been reported for TFLN PICs. Firstly, a bilayer inversely tapered mode size converter with a coupling loss of 1.7 dB/facet was demonstrated [31]. To further improve the mode match, polymer cladding waveguides were introduced to the bilayer couplers. The measured coupling loss is as low as 0.5 dB/facet, while facing challenges in mechanical and thermal stability [32]. Similarly, SiON was harnessed as the cladding material, and a coupling loss of 0.54 dB was observed for TE fundamental modes [33]. Moreover, silicon nitride assisted tri-layer edge couplers with 0.75 dB/facet loss were realized in 2023 [34]. Furthermore, a z-propagate cantilever-based polarization-diversity edge coupler with a 1.06 dB/facet was demonstrated [35]. However, in the above-mentioned designs, the minimum linewidth of the taper tips are quite narrow (typically <150 nm), requiring E-beam lithography (EBL) or deep ultraviolet lithography machine in the fabrication processes. Besides, multi-layer structures ask for precise alignment and complex manufacturing processes that are costly and time-consuming [36].

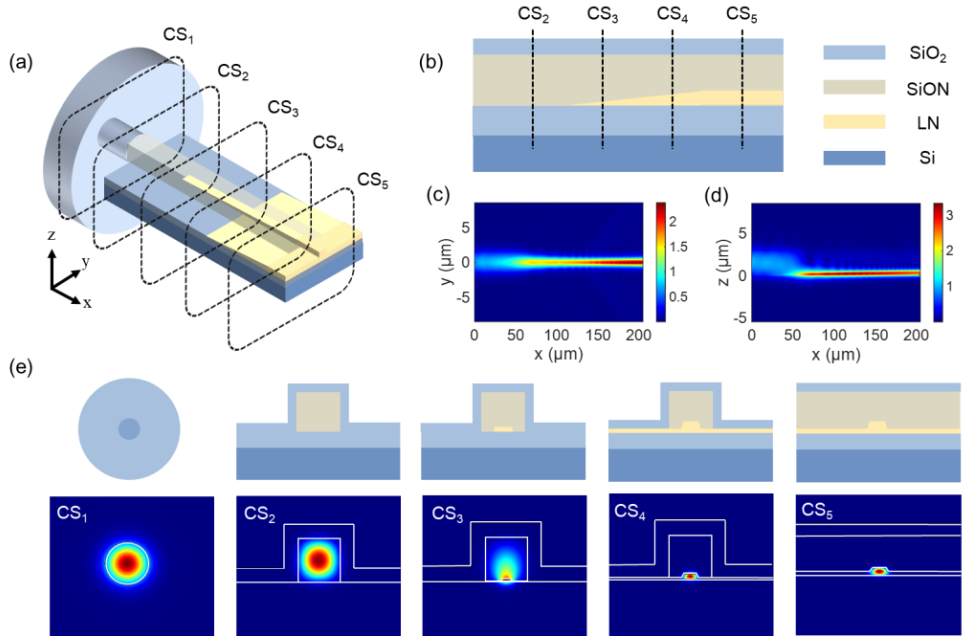

**Fig. 1.** (a) The schematic of the edge coupler. (b) The side view of the edge coupler. The simulated transmission of the E-field in the top (c) and side (d) views of the mode conversion near the wedge-shaped waveguide end. (e) Cross-section views of the coupler and corresponding field distribution of $TE_0$ mode.

To address this issue, we propose an efficient, broadband edge coupler with a TFLN waveguide with wedge-shaped ends and SiON waveguide covering the waveguide ends. When coupling with a 3-μm MFD lensed fiber, a coupling loss of 1.52 dB/facet is experimentally achieved at 1570 nm with a 3-dB bandwidth of over 90 nm. The edge coupler is also packaged with a UHNA7 fiber and displays an ultra-low loss of 0.92 dB/facet. It exhibits a relatively large tolerance for manufacturing misalignment and fiber-chip misalignment. The minimum width in the device is 600 nm, making its preparation compatible with the photolithography process and suitable for mass production at a relatively low cost.

## 2. DESIGN OF TFLN EDGE COUPLER

The scheme of the basic coupler structure is demonstrated in Fig. 1(a). The edge coupler consists of a wedge-shaped LN waveguide and a SiON waveguide. Primarily, light from the fiber couples into the SiON waveguide at the edge of the chip. Then, the wedge taper adiabatically converts the fundamental TE mode of the SiON waveguide into the LNOI waveguide mode by the gradual thickness variation. In contrast, many integrated TFLN devices rely on the double-layer rib geometries to convert the etched-rib waveguide mode into ridge TFLN waveguide modes [31-33]. In our work, the width gradient is replaced by a thickness gradient in the vertical dimension, avoiding the subwavelength-linewidth features and fabrication errors from aligning double-layer structures.

The side view of the edge coupler is shown in Fig. 1(b). The material index of SiON is optimized at 1.55, which is lower than the LN (2.21) and slightly higher than the silicon dioxide ($SiO_2$) slab (1.44), thus preventing light leakage into the slab modes. The simulated transmission of the electric (E) field in the top and side views of the mode conversion in the ramp waveguide are displayed in Fig. 1(c) and Fig. 1(d), respectively. Different cross-section views and the corresponding mode distributions along the edge coupler are shown in Fig. 1(e). The size of the SiON is optimized by calculating the mode overlap between the fiber mode and the SiON mode, as shown in the $CS_1$ and $CS_2$ profiles, respectively. As the ramp becomes gradually thicker, the mode is progressively bounded, as shown in $CS_3$ and $CS_4$. Finally, we get the normal LN ridge waveguide mode, as shown in $CS_5$.

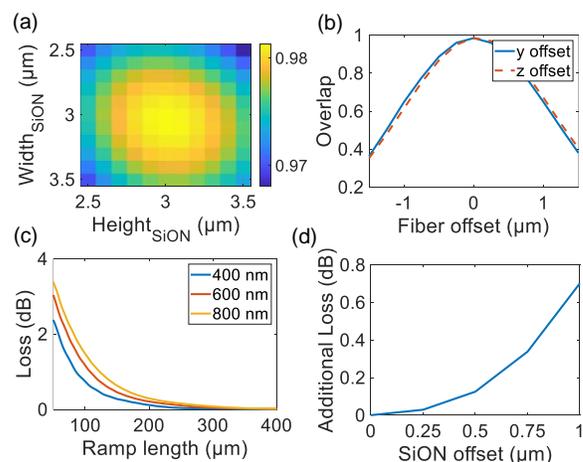

**Fig. 2** (a) Calculated overlap results between the fiber mode and the SiON waveguide mode for the waveguide width and height. (b) Tolerance for overlap of the mode field with the fiber offset. (c) Dependence of the transmission from SiON to TFLN waveguide on the ramped waveguide width and length. (d) Additional loss due to offset of SiON overlay from the TFLN waveguide.

The mode-overlap results with the width and height of SiON varying from 2.5 μm to 3.5 μm are given in Fig. 2(a). The optimal parameters are both chosen to be 3 μm to achieve a high coupling efficiency to fibers with an MFD of 3 μm. The results indicate the edge coupler is well tolerant to the size of the SiON waveguide cross-section. When the fiber deviates from the SiON waveguide, the mode-field overlap decreases, as shown in Fig. 2(b), exhibiting a decent fiber-offset tolerance. Due to the asymmetry of the material, the overlap drop in the Z direction is slightly asymmetric. The ramp width and length determine the coupling efficiency from SiON to LN ridge waveguides, as shown in Fig. 2(c). Because there is a certain tolerance in the ramp length by chemical-mechanical polishing (CMP) preparation, the impact of the ramp length on coupling loss is analyzed from a simulation perspective. With a length of around 200 μm, the coupling loss can drop to less than 0.3 dB. Moreover, only 0.45 dB extra loss is induced for over a 150 μm length offset (150 μm to 300 μm). A narrower ramped LN waveguide requires a shorter coupling length but is more demanding to fabrication. In our device, the width of the fabricated LN ramp waveguide is set to be 600 nm, which is compatible with the i-line photolithography process. The additional losses resulting from the SiON offset with the LN in the Y direction are shown in Fig. 2(d). The additional loss caused by the overlay manufacturing error of 1 μm is less than 0.7 dB. Considering the power coupling loss at the interface between fiber and chip as well as the mode conversion loss from $CS_2$ to $CS_5$, a total coupling loss of 0.43 dB/facet is obtained for $TE_0$ mode in simulations.

Experimentally, an X-cut TFLN with a 600-nm thick LN thin film sitting on top of a 2.5-μm thick $SiO_2$ layer on a Si substrate was used to fabricate the devices. The fabrication process is illustrated in Fig. 3(a). The size of the TFLN wafer is 1.2 cm × 1 cm. Firstly, we deposited 400-nm thick chromium (Cr) film in the middle of the TFLN wafer using a magnetron sputtering method. Using the deposited Cr as a hard mask, chemical mechanical polishing (CMP) was utilized to remove the LN film on the ends of the wafer while protecting the central area of the TFLN wafer from thinning. In the CMP process, a standard wafer polishing machine, polishing suspensions with silicon dioxide grains, and a soft velvet polishing cloth were harnessed [37]. The Cr mask and the exposed LN film near two edges contacted the polishing slurry. As Cr (Mohs 9) is much harder than LN (Mohs 5), LN film thins much faster than the Cr. After the CMP process, there is a 600-nm height difference between the edge area and the intermediate Cr covering area, connected by a 200 μm to 300 μm long ramp area, as shown in Fig. 3(b) and Fig. 3(c). For later fabrication, the sample was immersed in a Cr etching solution for 10 mins to remove the residual Cr mask.

Subsequently, electron-beam lithography (EBL) was used to define the waveguide pattern, and then followed by inductively coupled plasma reactive ion etching (ICP-RIE) to transfer it into the LN film and got the ridge waveguide with a 300-nm rib height and 60° sidewall angle. Here, the minimum width of our edge coupler is 600 nm, indicating that the EBL can be replaced by the ultra-violet (UV) photolithography process, which is less pricey and time-saving. Afterwards, the 3-μm thick SiON is deposited with plasma-enhanced chemical vapor deposition (PECVD). We patterned the SiON waveguide with AZ5214 resist by photolithography overlay process. After fully etched with an ICP-RIE process, the SiON waveguide extends along the LN ramp waveguide to the edge of the wafer. The microscope image of the ramped waveguides covered by the SiON waveguides is shown in Fig. 3(d). Then, a 3-μm thick silica cladding layer was deposited by PECVD to protect the SiON waveguide and improve the fiber alignment tolerances in later experimental tests. Finally, the chip facets were finely ground and polished by a series of sandpaper with a descending powder size. The polished edge cross-section is displayed in Fig. 3(e). The entire manufacturing process could be realized by UV photolithography other than EBL or deep UV, and the alignment tolerance of our devices is friendly and less prone to error.

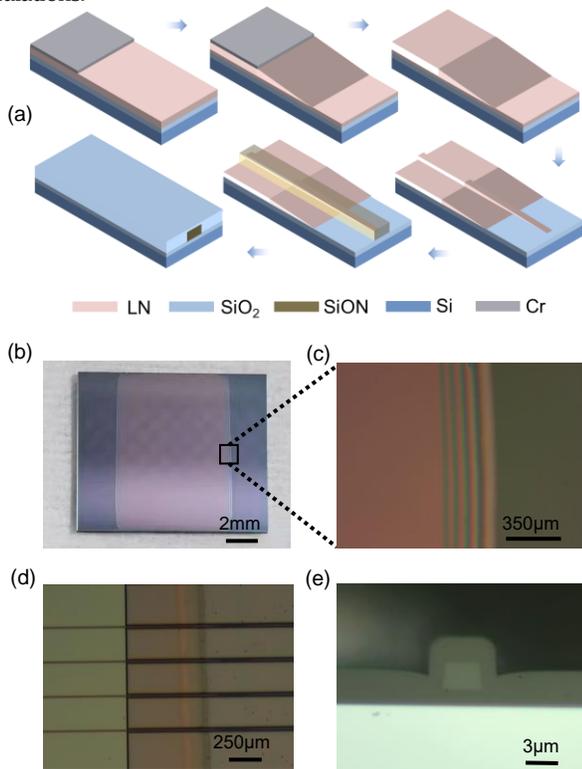

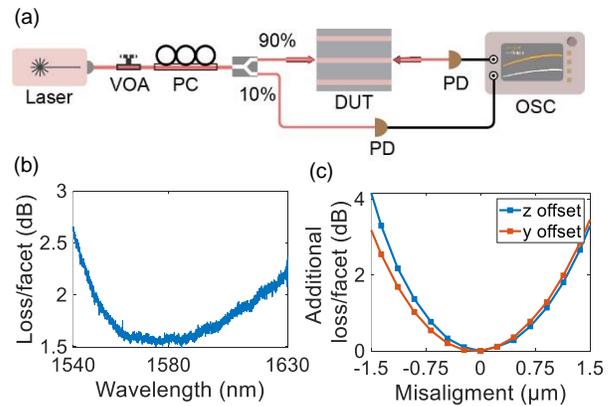

**Fig. 3.** (a) The fabrication process of the edge coupler. (b) The polished chip with two wedge-shaped edges. (c) The magnified microscopic view of the slope area. (d) The microscopic image of ramped waveguides covered by the SiON waveguides. (e) Microscopic view of polished SiON waveguide edge.

## 3. FABRICATION

**Fig. 4.** (a) The experimental setup to characterize the waveguide losses. (b) The experimentally measured coupling loss of the edge coupler from 1540 nm to 1630 nm. (c) The fiber alignment tolerance of the edge coupler.

## 4. EXPERIMENTAL SETUPS, RESULTS AND DISCUSSIONS

Figure 4(a) schematically illustrates the experimental setup to investigate the performance of the edge coupler. A tunable laser from 1510 nm to 1630 nm was used as the light source in the coupling test. After being attenuated by a variable optical attenuator (VOA) and polarized by a polarization controller (PC), the light was divided into two parts by a fiber coupler. The reference port (10%) was sent to a photodetector and collected to an oscilloscope to monitor the transmission spectrum of the pump. The signal port (90%) was coupled into the chip through a lensed fiber with an MFD of 3-μm. The output transmission spectrum was collected by another lensed fiber and pre-defined PD and then displayed on the oscilloscope through another channel.

The on-chip propagation loss was estimated by a ring resonator on the same chip, where the radiation loss can be ignored with a radius as large as 200-μm. The $Q$ factor at the 1550-nm band was measured by fitting the transmission spectrum. The loaded $Q$ factor is $2.86 \times 10^5$, and the intrinsic $Q$ factor is $5.07 \times 10^5$, corresponding to a propagation loss of 0.85 dB/cm. The broadband transmission of two edge couplers through a single waveguide was measured and normalized with the input pump, as illustrated in Fig. 4(b). Note that the length of the waveguide is 8 mm, ensuring passing through the slope area. The edge coupler keeps high coupling efficiency in a wide range over 90 nm from 1540 nm to 1630 nm, where the highest coupling efficiency is 1.52 dB/facet at 1570 nm. The broadband results were limited by the laser bandwidth. We can see that in the longer wavelength larger than 1630 nm, the edge coupler can still operate well. The main reason for the line dropping at short wavelengths is the mode mismatch between the SiON waveguide and the LN ramp waveguide. To further explore its potential in industrial applications, we also measured the coupling loss to the UHNA7 fiber (MFD 3.2μm). The refractive index liquid with a 1.47 index was used to minimize the reflective loss. It shows a loss with 0.92 dB/facet at 1570 nm, indicating that it could be an ideal candidate for packaging.

The fiber alignment tolerance was measured as well by shifting the fiber position in the Y direction and Z direction. As shown in Fig. 4(c), when the fiber misaligns 1.5 μm, less than 4 dB additional loss was measured, which means the edge coupler shows a good tolerance with both lateral and vertical deviation. Since the $SiO_2$ slab is only 2.5 μm thick, the light was induced to the higher-index silicon substrate (3.45). Therefore, there is a larger additional loss when the fiber goes down in the Z direction. A thicker slab, e.g. 4.7 μm, can prevent the light from being absorbed by the substrate, thereby decreasing the offset losses.

## 5. CONCLUSION

In conclusion, we demonstrate an edge coupler for TFLN waveguides, which is fabricated by cladding SiON waveguides on the wedge ends of TFLN waveguides. The proposed coupler shows high efficiency and robustness. A low coupling loss of 1.52 dB/facet and a wide operating band of over 90 nm has been achieved. A 0.92-dB/facet coupling loss was further observed between the edge coupler and a UHNA7 fiber. Both the manufacturing process and the optical misalignment have a large tolerance. The fabrication of the proposed coupler can be realized by an i-line stepper lithography machine, allowing for massive production at a low cost. This work opens the avenue of low-cost large-scale preparation of TFLN devices with low insertion loss.

**Funding.** This work was supported by the National Key Research and Development Program of China (Grant Nos. 2022YFA1404602), the National Natural Science Foundation of China (Grant Nos. 92250302, 12034010, 12134007, 92050111, and 12074199), and the 111 Project (Grant No. B23045).

**Disclosures.** The authors declare no conflicts of interest.

**Data availability.** Data underlying the results presented in this paper are not publicly available at this time but may be obtained from the authors upon reasonable request.